\documentstyle{elsartwb}
\input epsf

\begin{document}

\begin{frontmatter}

\hfill INP 1828/PH

\vspace{0mm}

\title{Gauging non-local quark models\thanksref{grants}}
\thanks[grants]{Research supported by
        the Polish State Committee for
Scientific Research grant 2P03B-080-12, and by the
Ministry of Science of Slovenia,
Scientific and Techological Cooperation Joint Project between
Poland and Slovenia}
\thanks[emails]{%
\hspace{0mm} broniows@solaris.ifj.edu.pl}
\author{Wojciech Broniowski}
\address{H. Niewodnicza\'nski Institute of Nuclear
Physics, PL-31342 Krak\'ow, Poland}

\begin{abstract}
Talk presented at the Mini-Worshop on {\em Hadrons as Solitons}, Bled, Slovenia,
6-17 July 1999
\end{abstract}

\end{frontmatter}

\section{Introduction and summary}

This research is being done together with Georges Ripka and Bojan Golli. We
will show how to gauge effective quark models with non-local quark
interactions. First, however, let me state the reasons why we want to use
such models:

\begin{itemize}
\item  Non-local regulators arise naturally in several approaches to
low-energy quark dynamics, such as the instanton-liquid model \cite
{Diakonov86} or Schwinger-Dyson calculations \cite{Roberts92}, presented at
this Workshop by Dubravko Klabu\v car. For the derivations and applications
of non-local quark models see, {\em e.g.},\cite
{Ripka97,Cahill87,Holdom89,Ball90,Krewald92,Ripka93,BowlerB,PlantB}. Hence,
we have to cope with non-localities from the outset.

\item  Non-local interactions regularize the theory in such a way that the
anomalies are preserved \cite{Ripka93,Arrio} and charges are properly
quantized. Recall that with other methods, such as the proper-time
regularization or the quark-loop momentum cut-off \cite{Goeke96,Reinhardt96}
the preservation of the anomalies can only be achieved if the (finite)
anomalous part of the action is left unregularized, and only the
non-anomalous part is regularized. If both parts are regularized, anomalies
are violated badly \cite{krs}. We consider such division rather artificial
and find it quite appealing that with non-local regulators both parts of the
action are treated on equal footing.

\item  With non-local interactions the effective action is finite to all
orders in the loop expansion. In particular, meson loops are finite and
there is no need to introduce another cut-off, as was necessary in the case
of local interactions \cite{Ripka96b,Tegen95,Temp}. As the result, the model
has more predictive power.

\item  As Bojan Golli, Georges Ripka and WB have shown \cite{nls}, stable
solitons exist in a chiral quark model with non-local interactions without
the extra constraint that forces the fields to lie on the chiral circle.
Such a constraint is external to the known derivations of effective quark
models and it is nice that we do not need it any more.
\end{itemize}

What is the price to pay for all these nice features?

\begin{itemize}
\item  The calculations are more complicated --- an extra integration over
an energy variable has to be performed numerically.

\item  Noether currents acquire non-local contributions. These are very much
wanted since they make the Noether currents and anomalies conserved.
However, there is an ambiguity involved. The transverse parts of currents
are not fixed and their choice is part of the model building. Recall that
this problem has been known for a long time in nuclear physics, where the
transverse part of the meson-exchange currents is ambiguous. It is not
possible to get rid of this ambiguity when one gauges non-local models. An
ideal solution would be to first gauge the underlying theory ({\em e.g}. QCD
with instantons) and then derive an effective gauged quark model. This has
not been attempted so far and we need to deal with transverse currents which
are not unique.

\item  Non-local interactions modify the current-quark interaction vertex.
In addition, contact terms (sea-gulls) are present in processes with more
than one current.
\end{itemize}

In this work we adopt the so-called P-exponent prescription \cite
{Bos,BowlerB,PlantB} for constructing Noether currents, with a focus on
application to solitons. In particular, we show that:

\begin{itemize}
\item  Noether currents corresponding to symmetries are conserved. In
particular, the CVC and PCAC relations hold.

\item  Charges of the soliton (baryon number, isospin, $g_A$) do not depend
on the choice of the path in the P-exponent, hence are unambiguously
determined. They pick up a non-local piece, which is crucial for the charge
quantization.

\item  Any n-point Green's function with external current momenta set to $0$
is independent of the path. An important example is the moment of inertia of
hedgehog solitons, or transverse parts of vector correlators at zero
momentum.

\item  Soliton radii, magnetic moments, in general form factors do depend on
the path, hence are not uniquely determined. We show that this dependence is
weak in the weak-nonlocality limit.

\item  A popular choice of the path in the P-exponent is the straight line 
\cite{BowlerB,PlantB}. We give explicit expressions for evaluating Noether
currents with this prescription which can be used in soliton calculations.
\end{itemize}

\section{The model}

We are concerned with the chiral quark model with non-local interactions,
such as discussed in the talks by Georges Ripka and Bojan Golli. The
Lagrangian is given by 
\begin{eqnarray}
L &=&\int (dp)\bar \psi (-p)(p\cdot \gamma -m)\psi (p)-\int (dp)(dp^{\prime
})\bar \psi (-p)r(p)\Phi ^a(p-p^{\prime })\Gamma _ar(p^{\prime })\psi
(p^{\prime })  \nonumber \\
&&\ -\frac{a^2}2\int (dp)\Phi ^a(-p)\Phi _a(p).  \label{Lp}
\end{eqnarray}
For the purpose of this work we find more convenient to work in the momentum
representation. We use the notation $(dp)=\int \frac{d^4p}{(2\pi )^4}$, {\em %
etc}. The fields $\psi $ describe the quark, $m$ is the current quark mass, $%
r(p)=r(p^2)$ is the regulator (local in the momentum space), and $\Phi
^a(p)=\int d^4xe^{-ip\cdot x}\Phi ^a(x)$ is the Fourier transform of the
soliton field, which is local in the coordinate space. The index $a=0$
corresponds to the $\sigma $ field, with $\Gamma _0=1$, and $a=1,2,3$
denotes the pion, with $\Gamma _a=i\gamma _5\tau _a$. We will also use the
abbreviation ${\bf \Phi }=\Phi ^a\Gamma _a$. The Euler-Lagrange equations
have the form

\begin{eqnarray}
(p\cdot \gamma -m)\ \psi (p) &=&\int (dp^{\prime })r(p)\Phi ^a(p-p^{\prime
})\Gamma _ar(p^{\prime })\psi (p^{\prime }),  \nonumber \\
\bar \psi (-p^{\prime })\ (p^{\prime }\cdot \gamma -m)\ &=&\int (dp)\bar
\psi (-p)r(p)\Phi ^a(p-p^{\prime })\Gamma _ar(p^{\prime }),  \label{eqp} \\
\Phi ^a(q) &=&-\frac 1{a^2}\int (dp^{\prime })\bar \psi (-q-p^{\prime
})r(q+p^{\prime })\Gamma ^ar(p^{\prime })\psi (p^{\prime }).  \nonumber
\end{eqnarray}
We note that by ``unbosonizing'' the model, {\em i.e.} by inserting the
third of equations (\ref{eqp}) into Eq. (\ref{Lp}), we recover the form with
the quartic separable quark interaction given by 
\begin{eqnarray}
L &=&\int (dp)\bar \psi (-p)(p\cdot \gamma -m)\psi (p)+\frac 1{2a^2}\int
(dp_1)(dp_2)(dp_3)(dp_4)\times  \nonumber \\
&&\delta (p_1+p_2+p_3+p_4)\bar \psi (p_1)r(p_1)\Gamma _ar(p_2)\psi (p_2)\bar
\psi (p_3)r(p_3)\Gamma _ar(p_4)\psi (p_4).  \label{fq}
\end{eqnarray}
On the other hand, integrating out the quark fields from Eq. (\ref{Lp})
leads to the bosonized (Euclidean) action, as used by George Ripka and Bojan
Golli in their talks: 
\begin{equation}
I(\Phi )=-{\rm Tr}\log \left( \partial _\tau -i\alpha \cdot \nabla +\beta
m+\beta r{\bf \Phi }r\right) +\frac 1{2a^2}{\rm Tr}\Phi ^2,  \label{Iac}
\end{equation}
where ${\rm Tr}$ denote the full trace,{\em \ i.e.} functional as well as
over color, flavor, and Dirac space in the case of quarks. The forms (\ref
{Lp}), (\ref{fq}) and (\ref{Iac}) are fully equivalent.

So much for the model. Its vacuum sector, spontaneous breaking of the chiral
symmetry, the method of fixing the model parameters by fitting the pion mass
and decay constant, and so on, are described, {\em e.g.}, in Refs. \cite
{BowlerB,PlantB,nls}.

\section{Gauge transformations}

The Noether construction of currents produces a contribution whenever a
derivative acts on a field in the Lagrangian. In Eq. (\ref{Lp}) the first
term (the local term) involves one derivative, and results in the usual
contribution to Noether currents. However, the interaction term with
functions $r(p_i)$ may be viewed, in the coordinate representation, as
involving infinitely many derivatives acting on the quark field. This leads
to complications. Below we show how to gauge the model in this case. Also,
the presence of infinitely many derivatives does not allow for the canonical
quantization of the quark field. We do not know how to quantize the quark
field, and yet, as we will show further on, the charges, such as the baryon
number, will be quantized.

Let us consider the gauge transformations of the quark field: 
\begin{equation}
\psi (x)\rightarrow e^{-i\lambda ^a\phi _a(x)}\psi (x),  \label{gau}
\end{equation}
where for the cases of interest $\lambda ^a=1/N_c$ (baryon current), $%
\lambda ^a=\tau ^a/2$ (isospin current), $\lambda ^a=\gamma _5\tau ^a/2$
(axial current). The phases $\phi _a(x)$ parameterize the transformation.
The local contribution to the Noether currents,{\em \ i.e.} the contribution
coming from the first term in Eq. (\ref{Lp}) is, of course, $j_a^{\mu ,{\rm L%
}}(x)=\bar \psi (x)\lambda ^a\gamma ^\mu \psi (x)$, or, in the momentum
representation, 
\begin{equation}
j_a^{\mu ,{\rm L}}(q)\equiv \int d^4x\ e^{-iq\cdot x}j_a^{\mu ,{\rm L}%
}(x)=\int (dp)\bar \psi (-p)\lambda _a\gamma ^\mu \psi (p+q).  \label{jlocq}
\end{equation}
With help of the equations of motion (\ref{eqp}) we find that 
\begin{equation}
q\cdot j_a^{{\rm L}}(q)=\int (dp)(dp^{\prime })\psi ^{\dagger }(-p^{\prime
})\left\{ 
\begin{array}{c}
r(p^{\prime }+q)\lambda ^a\beta {\bf \Phi }(p^{\prime }-p+q)r(p)- \\ 
r(p^{\prime })\beta {\bf \Phi }(p^{\prime }-p+q)\lambda ^ar(p-q)
\end{array}
\right\} \psi (p).  \label{qdjn}
\end{equation}
Note that $q\cdot j_a^{{\rm L}}(q)$ does not vanish in the non-local model!
What is missing is the non-local part discussed below.

\section{P-exponents}

Now we are going to adopt a rather elegant way of gauging the non-local
model. The P-exponent is defined as \cite{Bos,BowlerB,PlantB} 
\begin{equation}
W(x,y)=Pe^{i\int_x^y{\bf A}^\mu (s)ds_\mu },  \label{W}
\end{equation}
where ${\bf A}^\mu =\lambda ^aA_a^\mu $ is the gauge field (in general
non-abelian), $s$ parametrizes an (arbitrary) path from $x$ to $y$, and $P$
denotes the ordering along the path (needed only for non-abelian groups).
Under the gauge transformation the field $A_a^\mu $ transforms as $A_a^\mu
\rightarrow A_a^\mu +\partial ^\mu \phi _a$. The following object transforms
properly under the gauge transformation: 
\begin{equation}
\Gamma =-\int d^4x\int d^4y\int d^4z\ \psi ^{\dagger }(x)\langle
x|r|z\rangle W(x,z)\beta {\bf \Phi }(z)W(z,y)\langle z|r|y\rangle \psi (y).
\label{Gamma}
\end{equation}
Hence, when gauging the non-local model, we replace the interaction term in
Eq. (\ref{Lp}) with Eq. (\ref{Gamma}).

The non-local contribution to Noether currents, defined as $j_a^{\mu ,{\rm NL%
}}(q)\equiv \delta \Gamma /\delta A_\mu ^a(-q)|_{A=0}$ is equal to 
\begin{eqnarray}
j_a^{\mu ,{\rm NL}}(q) &=&-\int d^4x\ d^4y\ d^4z\int
(dk_1)(dk_2)(dl)(dp)(dp^{\prime })e^{-ik_1\cdot x+ik_2\cdot y+il\cdot
z+ip\cdot (x-z)+ip^{\prime }\cdot (z-y)}  \nonumber \\
&&\ \times i\psi ^{\dagger }(-k_1)r(p)\left\{ \int_x^ze^{-iq\cdot s}ds_\mu \
\lambda _a\beta {\bf \Phi }(l)+\int_z^ye^{-iq\cdot s}ds_\mu \ \beta {\bf %
\Phi }(l)\lambda _a\right\} r(p^{\prime })\psi (k_2).  \nonumber \\
&&  \label{jnl}
\end{eqnarray}
Note that this expression is not unique, which is manifest in the freedom of
choice of the path in the $s$ integration.

\section{Two tricks}

In the following we shall use the following obvious formulas:

\begin{eqnarray}
q^\mu \int_a^be^{-iq\cdot s}ds_\mu &=&\int_{q\cdot a}^{q\cdot b}e^{-iq\cdot
s}d(q\cdot s)=ie^{-iq\cdot b}-ie^{-iq\cdot a},  \label{trick1} \\
\int_a^bds_\mu &=&b_\mu -a_\mu .  \label{trick2}
\end{eqnarray}
The point here is that, clearly, the right-hand-sides of the above formulas
carry no information on the choice of the path joining the end-points $a$
and $b$. Formula (\ref{trick1}) appears when the longitudinal parts of
currents are involved, and formula (\ref{trick2}) occurs when the momentum
of the current is zero.

\section{Conservation of currents}

With help of Eq. (\ref{trick1}) we easily find the nonlocal contribution to
the divergence of currents:

\begin{eqnarray}
q\cdot j_a^{{\rm NL}}(q) &=&-\int (dp)(dp^{\prime })\psi ^{\dagger
}(-p^{\prime })\left\{ 
\begin{array}{c}
\lbrack r(p^{\prime }+q)-r(p^{\prime })]\lambda _a\beta {\bf \Phi }%
(p^{\prime }-p+q)r(p)- \\ 
r(p^{\prime })\beta {\bf \Phi }(p^{\prime }-p+q)\lambda _a[r(p-q)-r(p)]
\end{array}
\right\} \psi (p).  \nonumber \\
&&  \label{diva}
\end{eqnarray}
Combining it with the local piece (\ref{qdjn}) yields the divergence of the
total Noether current, $j_a(q)=j_a^{{\rm L}}(q)+j_a^{{\rm NL}}(q)$ : 
\begin{equation}
q\cdot j_a(q)=\int (dp)(dp^{\prime })\psi ^{\dagger }(-p^{\prime
})r(p^{\prime })[\lambda _a,\beta {\bf \Phi }(p^{\prime }-p+q)]r(p)\psi (p).
\label{conserve}
\end{equation}
For the baryon current this is immediately $0$. For the isospin and axial
currents we use the equations of motion for the $\Phi $ fields to obtain 
\begin{eqnarray}
\partial \cdot V_a(x) &=&-\frac 12\epsilon ^{abc}\Phi _b(x)\langle \psi
r|x\rangle \Gamma _c\langle x|r\psi \rangle =-\frac{a^2}2\epsilon ^{abc}\Phi
_b(x)\Phi _c(x)=0,  \label{CVC} \\
\partial \cdot A_a(x) &=&-\frac 12\langle \psi r|x\rangle \{\tau ^a\gamma
_5,(\Phi _0(x)+m+i\gamma _5\tau ^b\Phi _b(x))\}\langle x|r\psi \rangle
=ma^2\Phi _a(x).  \label{PCAC}
\end{eqnarray}
This verifies explicitly CVC and PCAC for the construction with P-exponents.
Note that the the above expressions hold for any choice of the path. In
other words, the longitudinal parts of vector and axial currents are fixed
unambiguously. We add parenthetically that this fact is related to the
Ward-Takahashi identities, which hold in the non-local models.

\section{Charges}

Equation (\ref{jnl}) simplifies greatly for the static case, $q=0$. Then,
through the use of Eq. (\ref{trick2}) we get 
\begin{eqnarray}
j_a^{\mu ,{\rm NL}}(0) &=&-\int d^4x\ d^4y\ d^4z\int
(dk_1)(dk_2)(dl)(dp)(dp^{\prime })\times  \label{Q1} \\
&&\ \ \left\{ \psi ^{\dagger }(-k_1)r(p){\bf \lambda }_a\beta {\bf \Phi }%
(l)r(p^{\prime })\psi (k_2)\frac \partial {\partial p_\mu ^{}}e^{-ik_1\cdot
x+ik_2\cdot y+il\cdot z+ip\cdot (x-z)+ip^{\prime }\cdot (z-y)}+\right. 
\nonumber \\
&&\ \ \left. \psi ^{\dagger }(-k_1)r(p)\beta {\bf \Phi }(l){\bf \lambda }%
_ar(p^{\prime })\psi (k_2)\frac \partial {\partial p_\mu ^{\prime
}}e^{-ik_1\cdot x+ik_2\cdot y+il\cdot z+ip\cdot (x-z)+ip^{\prime }\cdot
(z-y)}\right\}  \nonumber \\
&&  \nonumber
\end{eqnarray}
Next, we integrate by parts in the $p$ and $p^{\prime }$ variables, denote $%
r_\mu (p)\equiv dr_\mu (p)/dp^\mu $, and carry out the $x$, $y$, and $z$
integrations. The result is 
\begin{equation}
j_a^{\mu ,{\rm NL}}(0)=-\int (dp)(dp^{\prime })\psi ^{\dagger }(-p)\left\{ 
\begin{array}{c}
r_\mu (p){\bf \lambda }_a\beta {\bf \Phi }(p-p^{\prime })r(p^{\prime })+ \\ 
r(p)\beta {\bf \Phi }(p-p^{\prime }){\bf \lambda }_ar_\mu (p^{\prime })
\end{array}
\right\} \psi (p^{\prime }).  \label{Q2}
\end{equation}
When the quark fields are integrated out, the following formula for the full
currents at $q=0$ holds: 
\begin{equation}
j_a^\mu (0)=-{\rm Tr}\frac 1{\beta (-i\partial \cdot \gamma +r\Phi r)}\left(
\beta \gamma _\mu {\bf \lambda }_a+r_\mu {\bf \lambda }_a\beta \Phi r+r\beta
\Phi {\bf \lambda }_ar_\mu \right) .  \label{Q3}
\end{equation}
For the charges, $Q_a=j_a^0(0)$, we have 
\begin{equation}
Q_a=-{\rm Tr}\frac 1{\beta (-i\partial \cdot \gamma +r\Phi r)}\left( \beta
\gamma _0{\bf \lambda }_a+r_0{\bf \lambda }_a\beta \Phi r+r\beta \Phi {\bf %
\lambda }_ar_0\right) .  \label{Q4}
\end{equation}
As advocated earlier, $j_a^\mu (0),$ or the charges do not depend on the
path in the P-exponents. Certainly, this should make us happy. In local
field theories the charge is fixed by quantization. Here we can see a
similar feature, although we have not quantized the quark field.

\section{Baryon number of the soliton}

We shall now examine the issue of the baryon number in some greater detail.
For stationary solutions (such as solitons) the fields $\Phi $ are
time-independent. In this case (we pass to Euclidean space in this section) 
\begin{equation}
Q_a=-\sum_k\int_{-\infty }^\infty \frac{dv}{2\pi }\frac 1{i\nu +\varepsilon
_k\left( v\right) }\langle k,\nu |\left( \lambda _a-ir_0{\bf \lambda }%
_a\beta \Phi r-ir\beta \Phi {\bf \lambda }_ar_0\right) |k,\nu \rangle ,
\label{Q5}
\end{equation}
where we have used the spectral representation of the energy-dependent Dirac
Hamiltonian 
\begin{equation}
h(\nu )\equiv -i\vec \alpha \cdot \vec \nabla +m+\beta r(\nu )\Phi r(\nu ),
\label{dirham}
\end{equation}
with the energy-dependent spectrum: 
\begin{equation}
h(\nu )|k,\nu \rangle =\varepsilon _k\left( v\right) |k,\nu \rangle .
\label{spec}
\end{equation}
In particular, for the baryon number we obtain 
\begin{eqnarray}
B &=&-\frac 1{N_c}\sum_k\int_{-\infty }^\infty \frac{dv}{2\pi }\frac 1{i\nu
+\varepsilon _k\left( v\right) }[1-i\langle k,\nu |\beta (r_0\Phi r+r\Phi
r_0)|k,\nu \rangle ]  \nonumber  \label{Q6} \\
\ &=&-\frac 1{N_c}\sum_k\int_{-\infty }^\infty \frac{dv}{2\pi }\frac{%
1-i\langle k,\nu |h_0|k,\nu \rangle }{i\nu +\varepsilon _k\left( v\right) }%
=-\frac 1{N_c}\sum_k\int_{-\infty }^\infty \frac{dv}{2\pi i}\frac{1-i\frac{%
d\varepsilon _k\left( v\right) }{dv}}{\nu -i\varepsilon _k\left( v\right) },
\label{Q7}
\end{eqnarray}
where $h_0\equiv dh(\nu )/dv=\beta (r_0\Phi r+r\Phi r_0)$, and the
Feynman-Hellman theorem has been used in the last equality.

Suppose we have a pole in the quark propagator at $v=v_0$, {\em i.e.} 
\begin{equation}
\nu _0-i\varepsilon _k\left( v_0\right)  \label{valorb}
\end{equation}
($k$ labels quantum numbers relevant for the soliton, such as grand spin,
parity, and radial number). Expanding the denominator around $v=v_0$ we find 
\begin{eqnarray}
&&\ \nu _0+\left( v-v_0\right) -i\varepsilon _k\left( v_0\right) -i\left. 
\frac{d\varepsilon _k\left( v\right) }{dv}\right| _{v=v_0}\left(
v-v_0\right) +...  \label{denom} \\
\ &=&\left( v-v_0\right) \left( 1-i\left. \frac{d\varepsilon _k\left(
v\right) }{dv}\right| _{v=v_0}\right) +...  \nonumber \\
&&  \nonumber
\end{eqnarray}
We notice that, quite remarkably, the numerator in (\ref{Q7}) is such, that
the residue of any pole is equal to unity. This means, that the baryon
number is properly quantized in the model.{\em \ }Therefore, we have
achieved the baryon quantization without quantizing the quark field! This is
a very important feature, which brings the model close to the particle-hole
interpretation: by occupying ($N_c$ times) a pole of the quark propagator we
raise the baryon number by one unit. We achieve this by distorting the
contour in the $v$ integration such that it encircles the occupied valence
states (this is only necessary if these states lie above the real axis). See
Fig. 1.

It can also be shown straightforwardly that the energy of the stationary
system equals to 
\begin{equation}
E=-\sum_k\int_{-\infty }^\infty \frac{dv}{2\pi i}\varepsilon _k\left(
v\right) \frac{1-i\frac{d\varepsilon _k\left( v\right) }{dv}}{\nu
-i\varepsilon _k\left( v\right) },  \label{En}
\end{equation}
hence occupying ($N_c$ times) a state corresponding to a pole at $v=v_0$
brings the energy $N_c\varepsilon _k\left( v_0\right) .$ The Dirac sea
contribution is obtained from expression analogous to (\ref{Q7}), but with
the contour undistorted.

Above we have said ``{\em close }to the particle-hole interpretation'' for
the following reason. Unlike the usual many-body problem, where all poles of
the quark propagator lie on the imaginary axis (recall we live in the
Euclidean space), in the non-local case there are in general many poles in
the complex plane. In fact, this feature is also present in certain local
quark models \cite{analyt}. These poles are induced by the presence of the
regulator. Since they are complex, they do form asymptotic states. In fact,
it is possible to choose the regulator in such a way, that in the vacuum
there are no poles on the imaginary axis (no ``physical'' poles), which is
sometimes referred to as ``analytic confinement''. Remarkably, the hedgehog
soliton fields generate a valence state on the imaginary axis with a low
eigenvalue (Bojan Golli's talk). It is therefore natural to occupy this
state, thus providing the soliton a unit of baryon number.

\begin{figure}[t]
\vspace{0mm} \epsfxsize = 6.5 cm \centerline{\epsfbox{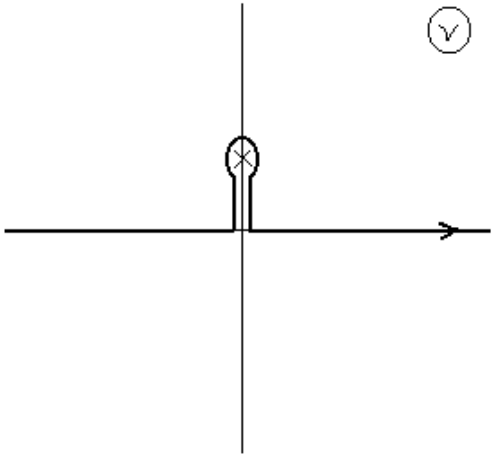}} \vspace{0mm} 
\label{diag}
\caption{The contour in the integral over $\nu$ encircling a positive-energy
valence state.}
\end{figure}

The contribution of the Dirac sea levels to observables ({\em e.g.} as in (%
\ref{En})) is obtained by carrying numerically the integration over the $v$
variable along the real axis.

We also remark that expressions analogous to (\ref{Q7}) hold for any
``good'' quantum number, {\em i.e.} when $[h\left( \nu \right) ,\lambda
_a]=0 $. For instance, in the Friedberg-Lee-like model, where $\Phi =\sigma $
(no pion field present), the isospin is a good quantum number, and we have,
in analogy to (\ref{Q7}), 
\begin{equation}
I_3=-\frac 12\sum_k\int_{-\infty }^\infty \frac{dv}{2\pi i}\frac{(1-i\frac{%
d\varepsilon _k\left( v\right) }{dv})\langle k,\nu |\tau _3|k,\nu \rangle }{%
\nu -i\varepsilon _k\left( v\right) },\quad ({\rm Freedberg-Lee})  \label{Q8}
\end{equation}
In hedgehog models isospin is not a good quantum number and it has to be
restored by a suitable projection method, {\em e.g.} by cranking \cite
{ANW83,CB86}.

\section{$g_A$}

Another important quantity is the axial charge of the nucleon, $g_A$,
evaluated in hedgehog models from the expectation value of the $z$ component
of the axial current. In fact, we see that also the space components of
current do not depend on the path at $q=0$. Hence $g_A$\ is
path-independent. An explicit expression can be immediately derived from (%
\ref{Q3}), using the method of Ref. \cite{CB86}.

\section{Moment of inertia}

It is well-known that hedgehog solitons break the spin and isospin, which
are restored by a suitable projection method. In the {\em cranking} method 
\cite{ANW83,CB86} the basic dynamical quantity is the moment of inertia $%
\theta $. It is obtained by adiabatically rotating the soliton, 
\begin{equation}
\psi (x)\rightarrow e^{-i\tau ^a\Omega _at}\psi (x),  \label{cra}
\end{equation}
where $\Omega _a$ is the (adiabatically small) velocity of rotation in the
isospin space, and $t$ is the time. The moment if inertia is obtained by
performing the transformation (\ref{cra}) in the action and then identifying
the coefficient of $\Omega ^2.$ We notice that Eq. (\ref{cra}) is a special
case of gauge transformation (\ref{gau}), with the vector potential equal to 
$A_a^\mu =(\Omega _a,0,0,0)$. Therefore we should apply the prescription (%
\ref{Gamma}), and then evaluate $\theta $ from the formula $\delta
^2I/\delta \Omega _a\delta \Omega _b=\delta ^{ab}\theta $. Applying the same
techniques as used in the previous sections leads to the following
expression for the moment of inertia in models with non-local regulators
(Euclidean notation):

\begin{eqnarray}
\theta &=&-\frac 12\sum_{k,l}\int \frac{d\nu }{2\pi }\frac{\langle k,\nu
|\left( \tau _3-i\tau _3\beta r_0{\bf \Phi }r-i\beta r{\bf \Phi }r_0\tau
_3\right) |l,\nu \rangle }{i\nu +\varepsilon _k(\nu )}\times  \nonumber \\
&&\times \frac{\langle l,\nu |\left( \tau _3-i\tau _3\beta r_0{\bf \Phi }%
r-i\beta r{\bf \Phi }r_0\tau _3\right) |k,\nu \rangle }{i\nu +\varepsilon
_l(\nu )}  \nonumber \\
&&\ -\sum_k\int \frac{d\nu }{2\pi }\frac{\langle k,\nu |\beta \left( \frac
12r_{00}{\bf \Phi }r+\tau _3r_0{\bf \Phi }r_0\tau _3+\frac 12r{\bf \Phi }%
r_{00}\right) |k,\nu \rangle }{i\nu +\varepsilon _k(\nu )},  \label{mominer2}
\end{eqnarray}
where $r_{00}=d^2r/d\nu ^2$. The first term in Eq. (\ref{mominer2}) is
dispersive, it is involves to quark propagators. The second one is the
contact term (sea-gull), with one quark propagator looped around. In the
local limit Eq. (\ref{mominer2}) reduces formally to the usual Inglis
formula 
\begin{equation}
\theta =\frac 12\sum_{p,h}\frac{|\langle h|\tau _3|p\rangle |^2}{\varepsilon
_p-\varepsilon _h}.\qquad ({\rm local})  \label{Inglis}
\end{equation}
In the presence of valence states we proceed as in the case of the baryon
number or the soliton energy, and decompose the total moment of inertia into
the valence and Dirac sea parts. The valence part is obtained by explicitly
occupying the state satisfying Eq. (\ref{valorb}).

Notice that $\theta $ in Eq. (\ref{mominer2}) is path-independent. In fact,
it is clear from the derivation that any n-point Green's function with
vanishing momenta on the external current lines does not depend on the
choice of the path. This is because the differentiation of the action with
respect to the $A$ field at $q=0$ brings down, according to Eq. (\ref{trick2}%
), the factor of $(y-x)_\mu $, where $y$ and $x$ are end-points of the line.
Then $A$ is set to zero. Obviously, no information of the choice of the path
is left.

\section{Straight-line paths}

The quantities discussed above did not depend on the choice of the path.
This is not true for other physical quantities, such as form factors and
magnetic moments of baryons, or transverse vector correlators. The simplest
choice of the path in the P-exponent is just the straight line, as used in
Refs. \cite{BowlerB,PlantB}. One parameterizes $s^\mu =x^\mu +\alpha (z^\mu
-x^\mu )$. Then 
\begin{equation}
\int_x^ze^{-iq\cdot s}ds_\mu =\int_0^1d\alpha (z^\mu -x^\mu )e^{-iq\cdot
(x+\alpha (z-x))},  \label{straight}
\end{equation}
and the repetition of the steps of Eq. (\ref{Q1}-\ref{Q2}) leads to 
\begin{eqnarray}
j_a^{\mu ,{\rm NL}}(q) &=&-\int (dp)(dp^{\prime })\int_0^1d\alpha \psi
^{\dagger }(-p)\times  \label{jmunl} \\
&&\ \ \ \left[ 
\begin{array}{c}
r_\mu (p+\alpha q)\lambda _a\beta {\bf \Phi }(p-p^{\prime }+q)r(p^{\prime })+
\\ 
r(p)\beta {\bf \Phi }(p-p^{\prime }+q)\lambda _ar_\mu (p^{\prime }-\alpha q)
\end{array}
\right] \psi (p^{\prime }).  \nonumber
\end{eqnarray}
In the coordinate representation we can write equivalently 
\begin{eqnarray}
&&j_a^{\mu ,{\rm NL}}(x)=-\int_0^1d\alpha \int dz\ dx^{\prime }\delta
(x-x^{\prime }-\alpha (z-x^{\prime }))\times  \nonumber \\
&&\left[ \langle \psi |z\rangle \langle z|r_\mu |x^{\prime }\rangle \langle
x^{\prime }|\lambda _a\beta {\bf \Phi }r|\psi \rangle +\langle \psi |r\beta 
{\bf \Phi }\lambda _a|x^{\prime }\rangle \langle x^{\prime }|r_\mu |z\rangle
\langle z|\psi \rangle \right] .  \label{wnlx}
\end{eqnarray}
Expression (\ref{jmunl},\ref{wnlx}) can be used to calculate the non-local
contribution to currents.

\section{ Form factors}

Rewriting the general expression (\ref{jmunl}) for ${\bf \lambda }_a=1/N_c$
we find the following expression for the non-local contribution to the
Fourier-transformed baryon density: 
\begin{equation}
\rho ^{{\rm NL}}(q)=-\frac 1{N_c}\int (dp)(dp^{\prime })\int_0^1d\alpha \bar
\psi (-p)\left\{ 
\begin{array}{c}
r_0(p+\alpha q){\bf \Phi }(p-p^{\prime }+q)r(p^{\prime })+ \\ 
r(p){\bf \Phi }(p-p^{\prime }+q)r_0(p^{\prime }-\alpha q)
\end{array}
\right\} \psi (p^{\prime }).  \label{rhod}
\end{equation}
We can now pass to the Breit frame ($q_0=0$) and expand Eq. (\ref{rhod}) at $%
\vec q=0$. The term quadratic in $\vec q$ is related to the non-local
contribution to the baryon mean squared radius, which equals to 
\begin{eqnarray}
\langle r^2\rangle ^{{\rm NL}} &=&\frac 1{N_c}\sum_{i=1}^3\int
(dp)(dp^{\prime })\bar \psi (-p)\times  \label{rho2} \\
&&\ \left[ 
\begin{array}{c}
\left\{ r_0(p){\bf \Phi }_{ii}(p-p^{\prime })+r_{0i}(p){\bf \Phi }%
_i(p-p^{\prime })+\frac 13r_{0ii}(p){\bf \Phi }(p-p^{\prime })\right\}
r(p^{\prime })+ \\ 
r(p)\left\{ {\bf \Phi }_{ii}(p-p^{\prime })r_0(p^{\prime })-{\bf \Phi }%
_i(p-p^{\prime })r_{0i}(p^{\prime })+\frac 13{\bf \Phi }(p-p^{\prime
})r_{0ii}(p^{\prime })\right\}
\end{array}
\right] \psi (p^{\prime }).  \nonumber
\end{eqnarray}

The expressions for the isoscalar and isovector magnetic moment involve
cranking and are somewhat more complicated. They can be obtained using the
methods described above along the lines of Ref. \cite{CB86}.

\section{The weak-non-locality limit}

It is instructive to have a closer look at Eq. (\ref{rho2}). Suppose the
soliton has a typical size $R\sim 1/\Lambda _S$, and the regulator has a
momentum scale $\Lambda $ above which the momenta are suppressed, {\em e.g. }%
$r(p^2)=\exp (-p^2/\Lambda ^2)$. Derivatives with respect to momenta, such
as appear in Eqs. (\ref{rho2}), bring down a factor of the inverse scale
squared, {\em e.g. }$r_\mu (p^2)=-2p_\mu \exp (-p^2/\Lambda ^2)/\Lambda ^2$,
and similarly for the soliton profile, where $\Phi _j(P-P^{\prime
})=2(P-P^{\prime })_j/\Lambda _S^2$ $\,\Phi ^{\prime }(P-P^{\prime })$. In
the weak-non-locality limit,{\em \ i.e. }when $\Lambda \gg \Lambda _S$, the
terms with derivatives of the regulators are suppressed relative to the
terms with derivatives of $\Phi $. For instance, in Eq. (\ref{rho2}) we must
keep only the terms with $\Phi _{ii}$, and can neglect the remaining pieces.
In the coordinate representation this is equivalent to using, instead of (%
\ref{wnlx}), the following expression for the non-local current: 
\begin{equation}
\ j_a^{\mu ,{\rm NL}}(x)=-\ \left[ \langle \psi |x\rangle \langle x|r_\mu
\lambda _a\beta {\bf \Phi }r|\psi \rangle +\langle \psi |r\beta {\bf \Phi }%
\lambda _ar_\mu |x\rangle \langle x|\psi \rangle \right] .  \label{wnlxp}
\end{equation}
One can formally pass from Eq. (\ref{wnlx}) to Eq. (\ref{wnlxp}) by
commuting the $r_\mu $ and $|x\rangle \langle x|$ operators, which is
allowed in the weak-non-locality limit \cite{Ripka97}.

For the solitons shown by Bojan Golli $\Lambda \sim 1{\rm GeV}$ and $\Lambda
_S\sim .25{\rm GeV}$, hence $\Lambda _S^2/\Lambda ^2\sim 0.06$, hence we
seem to be very close to the week-non-locality limit. This is fortunate,
since then the observables such as the baryon radius, {\em etc.}, do not
depend strongly on the choice of the path in the P-exponent.

Our research on solitons with non-local regulators is under way and we hope
to be able present further exciting results shortly.

\ack
I am grateful to Nikos Stefanis, Maxim Polyakov and Enrique Ruiz Arriola for
many useful discussions on topics related to this talk. I wish to cordially
thank the organizers for their hospitality and for creating the perfect
working atmosphere at the Bled workshop.

\bibliographystyle{npa96}
\bibliography{wb,njl}

\end{document}